\documentclass[a4paper,fleqn]{cas-sc}
\usepackage{lineno}

\usepackage[numbers]{natbib}
\def\tsc#1{\csdef{#1}{\textsc{\lowercase{#1}}\xspace}}
\newcommand{\RNum}[1]{\uppercase\expandafter{\romannumeral #1\relax}}
\tsc{WGM}
\tsc{QE}

\begin{document}
\let\WriteBookmarks\relax
\def\floatpagepagefraction{1}
\def\textpagefraction{.001}

\shorttitle{Towards a new generation of solid total-energy detectors for neutron-capture time-of-flight experiments with intense neutron beams}    
\shortauthors{J. Balibrea-Correa and the n\_TOF collaboration}  


\title [mode = title]{Towards a new generation of solid total-energy detectors for neutron-capture time-of-flight experiments with intense neutron beams}

\author[1]{ J.~Balibrea-Correa} %
\ead{javier.balibrea@ific.uv.es}
\cortext[cor1]{Corresponding author}
\author[98,14]{V.~Babiano-Suarez} %
\author[1]{J.~Lerendegui-Marco} %
\author[1]{C.~Domingo-Pardo} %
\author[1]{I.~Ladarescu} %
\author[1]{A.~Tarife\~{n}o-Saldivia} %
\author[1]{G.~de~la~Fuente-Rosales} %
\author[1]{B.~Gameiro}
\author[99]{N.~Zaitseva}
\author[3]{V.~Alcayne} %
\author[3]{D.~Cano-Ott} %
% ...
%
\address[1]{Instituto de F\'{\i}sica Corpuscular, CSIC - Universidad de Valencia, Spain}
\address[3]{Centro de Investigaciones Energ\'{e}ticas Medioambientales y Tecnol\'{o}gicas (CIEMAT), Spain} %
\address[14]{Universitat Polit\`{e}cnica de Catalunya, Spain} %
% ...
\address[99]{Lawrence Livermore National Laboratory, 7000 East Avenue, Livermore, CA 94551, USA}
\address[98]{Universidad de Valencia, Spain} %

\begin{abstract}
Challenging neutron-capture cross-section measurements of small cross sections and samples with a very limited number of atoms require high-flux time-of-flight facilities. In turn, such facilities need  innovative detection setups that are fast, have low sensitivity to neutrons, can quickly recover from the so-called $\gamma$-flash, and offer the highest possible detection sensitivity. In this paper, we present several steps toward such advanced systems. Specifically, we describe the performance of a high-sensitivity experimental setup at CERN n\_TOF EAR2. It consists of nine sTED detector modules in a compact cylindrical configuration, two conventional used large-volume C$_{6}$D$_{6}$ detectors, and one LaCl$_{3}$(Ce) detector. The performance of these detection systems is compared using $^{93}$Nb($n$,$\gamma$) data. We also developed a detailed \textsc{Geant4} Monte Carlo model of the experimental EAR2 setup, which allows for a better understanding of the detector features, including their efficiency determination. This Monte Carlo model has been used for further optimization, thus leading to a new conceptual design of a $\gamma$ detector array, STAR, based on a deuterated-stilbene crystal array. Finally, the suitability of deuterated-stilbene crystals for the future STAR array is investigaged experimentally utilizing a small stilbene-d12 prototype. The results suggest a similar or superior performance of STAR with respect to other setups based on liquid-scintillators, and allow for additional features such as neutron-gamma discrimination and a higher level of customization capability.
\end{abstract}


\begin{highlights}
\item A hybrid detection system for high sensitivity and high accuracy neutron capture cross section measurements at CERN n\_TOF EAR2 is presented.
\item A versitile Monte-Carlo corresponding to the experimental setup is implemented, thereby enabling future enhancements of the experimental setup based on scintillation materials of new generation.
\item A conceptual design for a novel detection system based on crystals of deuterated stilbene is presented.
\item Experimental results utilizing a small deuterated stilbene prototype are reported, thereby addressing its main features for their array detector future implementation in n\_TOF experiments.
\end{highlights}



\maketitle





\clearpage 






\section{Introduction}\label{Sec:Introduction}
Historically, innovation on detection techniques and advances in high-quality pulsed neutron beams have led to fascinating discoveries in stellar nucleosynthesis and to subsequent refinements of theoretical models of stellar structure and galactic-chemical evolution \cite{Alpher:1948,Alpher:1948.1,Alpher:1948.2,Alpher:1948.3,Burbidge:1957,Kappeler:2011,Arcones:2022,Massimi:2022}. 
For ($n,\gamma$) cross section measurements in the energy range relevant for astrophysics scintillation detectors based on deuterated benzene (C$_{6}$D$_{6}$) are being extensively used at the most active pulsed time-of-flight facilities in the world, like CERN n\_TOF (Switzerland)~\cite{Mengoni:2019}, JRC (Belgium)~\cite{BENSUSSAN:1978} and Back-n (China)~\cite{Tang:2021}. To a large extent, the success of these liquid scintillators is based on the possibility of applying the pulse-height weighting technique (PHWT) to convert them into total-energy detectors (TED) in combination with their low neutron sensitivity~\cite{MacKlin:1967,Abbondanno:2004,Borella:2007,SCHILLEBEECKX:2012,Ren:2019}. The PHWT technique enables systematic uncertainties of only a few percent in the cross section determination, compatible with the accuracy required by stellar models~\cite{Kappeler:2011}.
In practice, the technique can be applied to any kind of $\gamma$-ray detector, provided that the detection probability remains low enough to avoid registering two- or more gamma-quanta from the same neutron
capture cascade. PHWT allows for large flexibility in the design of the detection apparatus and allows further system optimization for specific experimental conditions. In the past, these optimizations included the replacement of the first generation of TED detectors, C$_{6}$F$_{6}$~\cite{Corvi:1988}, by C$_{6}$D$_{6}$ as well as original housing and structural materials with carbon fibres to reduce the neutron sensitivity of the setup \cite{Plag:2003,Mastinu:2013}. Still, until recently most neutron-capture experiments have utilized liquid scintillators of relatively large volume ($\approx$500 mL up to 1 L). However, large detection volumes in close geometry are not suitable for high-flux neutron beams, such as the one at n\_TOF EAR2~\cite{Weiss:2015}. In particular, the detectors suffer from significant pile-up and dead-time effects due to a very high count rate that is difficult to manage~\cite{Balibrea:2024}. These problems then require larger distance to the sample, severely limiting the signal-to-background ratio, and thus affecting the detection sensitivity. In order to deal with such limitations, the segmented total-energy detector sTED~\cite{Alcayne:2023,Alcayne:2024} was designed for coping with the extreme conditions, reducing the active volume of individual detector cells in exchange of dealing with the dead-time corrections~\cite{Balibrea:2024}. 

The reduction in active volume of the $\gamma$-ray detectors has led the n\_TOF collaboration to devote significant efforts in recent years to developing new experimental setups aimed at increasing the sensitivity of the experiments while minimizing systematic errors that could affect the measurements. It is worth mentioning that in the framework of the n\_TOF collaboration, other research lines try to cope with other scenarios where high count rate is not an issue, demonstrating the possibility of combining the TED principle with $\gamma$-ray imaging techniques for background suppression \cite{Babiano:2021,Domingo:2016}. Additional studies have also shown the possibility of using the PHWT with high-efficiency set-ups, which aim to reduce statistical uncertainties and lower angular distribution effects \cite{Mendoza:2023}.
In this work, we report on the features of a high-sensitivity, state-of-the-art detection system for ($n$,$\gamma$) cross-section measurements at n\_TOF EAR2. The system consists of nine sTED detectors in a close cylindrical geometric configuration. In addition, two C$_{6}$D$_{6}$ detectors and one LaCl$_3$(Ce) detector are positioned at 135$^{\circ}$ and at greater distances from the sample. In practice, C$_{6}$D$_{6}$ presents significant disadvantages: it is highly toxic, flammable, and carcinogenic, thus requiring careful handling and costly maintenance. Therefore its replacement with a new material that would maintain, or even improve the overall detector performance appears to be necessary. One of the most promising options is the use of high-purity deuterated stilbene crystals \cite{Becchetti:2017,Carman:2018}, which, when combined with the high-sensitivity detector arrangement, could serve as a replacement for C$_{6}$D$_{6}$ detectors in the near future, as will be discussed in the second part of the manuscript.

The present article is structured as follows: Sec.~\ref{Sec:Stat} introduces briefly the concept of signal-to-background ratio and highlights the importance of this ratio for a ($n$,$\gamma$) experiment and nuclear data evaluations. Sec.~\ref{Sec:PoC} presents the 2022 state-of-the-art experimental setup for ($n$,$\gamma$) cross section measurements at CERN n\_TOF EAR2, specially designed to improve the sensitivity  with respect to previous setups. Sec.~\ref{Sec:MC} introduces a Monte Carlo (MC) model of the presented setup based on \textsc{Geant4}~\cite{Allison:2016} that allows one to quantify the efficiency of the detectors and serves as a starting point for tests of additional changes in the detection setup. Sec.~\ref{Sec:Stilbene} envisions a new generation of stilbene-d12 based total-energy detectors for future experiments at n\_TOF EAR2 and Sec.~\ref{sec:ExpStilbene} describes initial characterization for a small stilbene-d12 prototype.  Finally, Sec.~\ref{Sec:Conclusions} presents the main conclusions and outlook of this work.

\section{The importance of signal-to-background ratio for an experiment}\label{Sec:Stat}

In practical terms, the quality of a Time-of-Flight (ToF) based neutron capture cross-section experiment strongly depends on several ingredients, such as high-quality sample preparation~\cite{Chiera:2022}, beam-time restrictions, and strict control of systematic uncertainties associated with background determination and normalization~\cite{Lerendegui:2018,Balibrea:2020,Guerrero:2020}. However, as with any counting experiment the detection process is governed by statistics~\cite{Avanzi:2021}, and the actual number of detected events is thus influenced by statistical fluctuations. In addition, physical events are often detected together with other non-negligible background sources. One critical aspect in the design of the experiment thus involves optimizing the signal-to-background ratio ($s/b$) of the experimental setup. In practical applications, this ratio is defined as the number of registered counts during the measurement ($n_{i}$) divided by the number of background counts registered or estimated in the same region ($n_{b}$), normalized by the number of neutron pulses devoted to each configuration, $N^{\prime}$ and $N$, respectively. 
\begin{equation}
s/b = \frac{n_{i}}{n_{b} (N^{\prime}/N)} \equiv \frac{n_{i}}{n_{b} \alpha}
\end{equation}

In the case that the individual pulses would have different intensity, the normalization ($N^{\prime}$ and $N$) can be extended to the sum of the neutrons devoted for each configuration. Depending on the $s/b$ value, a larger or smaller number of neutron pulses is required to observe a significant result. The $s/b$ is closely related to the precision determining a neutron capture cross section and the attainable accuracy of the resonance parameters. In practice, for a counting experiment --where both $n_i$ and $n_b$ come from the Poisson distribution-- the statistical significance for an existence of a signal above the background can be described~\cite{Li:1983,Vianello:2018} by standard statistical $p$-value that is given by
\begin{equation}
   p\textrm{-value}\equiv\int^{\infty}_{\lambda}{N(x)dx} 
   \label{eq:p-value}
\end{equation}
where $N(x)$ is the normal distribution with zero mean and unit variance and  $\lambda$ 

\begin{eqnarray}
\lambda=\sqrt{2}\left[n_{i}\log{\left(\frac{\alpha+1}{\alpha}\frac{n_{i}}{n_{i}+n_{b}}\right)}
+n_{b}\log{\left(\left(\alpha+1\right)\frac{n_{b}}{n_{i}+n_{b}}\right)}\right]^{1/2}.
    \label{eq:S1}
\end{eqnarray}

Given a significance level, the $p$-value can be use to assess whether the observed data correspond to a background fluctuation or a non-zero cross section value. It is easy to check that the larger $s/b$, the higher the $p$-value and the more precise cross section or resonance parameters determination for a determined statistic level. The value of $s/b$ is particularly relevant for the challenging measurements of radioactive samples with a very small number of atoms available or samples embedded in matrix materials with large contributions from other isotopes~\cite{LerendeguiCDS:2020,BalibreaCDS:2020}. In addition, the specific characteristics of the ToF facility, such as instantaneous flux and resolution function, indirectly affect the $s/b$ and must be taken into account for the sensitivity of the experimental setups.

It is important to highlight at this point that evaluated libraries are built from experimental data and, therefore, uncertainties of the original works have a sizable impact on nuclear evaluations and any calculation derived from them either for nuclear astrophysics or applications~\cite{SCHILLEBEECKX:2012}. Thus, high sensitivity setups are required to improve the state-of-the-art of neutron induced cross sections, and this is in particular the case for the ($n$,$\gamma$) reaction channel. In the next sections we will report on our effort made in the direction of getting better $s/b$ compared to traditionally used C$_{6}$D$_{6}$ detection setup.

\section{Hybrid experimental setup for ($n$,$\gamma$) cross section measurements at CERN n\_TOF EAR2}\label{Sec:PoC}

\begin{figure*}
    \centering
    \begin{tabular}{c c}
    \includegraphics[width=0.5\columnwidth]{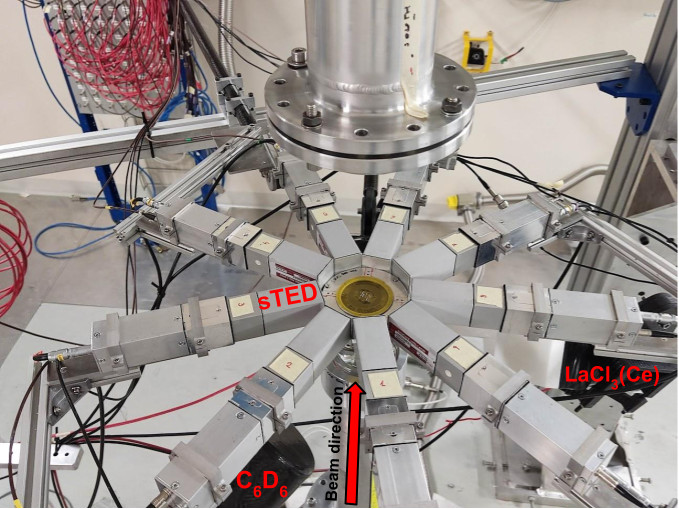} &
    \includegraphics[width=0.5\columnwidth]{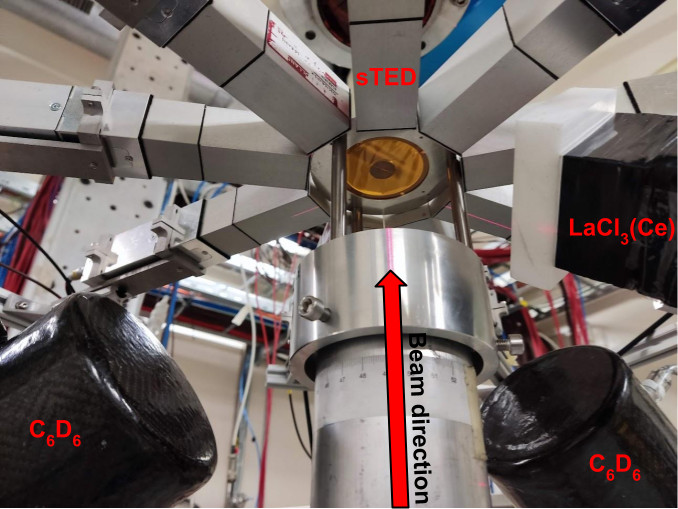} \\
    \end{tabular}
    \caption{Pictures of the hybrid ($n$,$\gamma$) experimental setup at EAR2 during the 2022 n\_TOF experimental campaign. The sTED cylindrical configuration surrounds the sample with cylindrical geometry. Two large-volume C$_{6}$D$_{6}$ detectors are placed in the upstream position, pointing toward the sample at 135$^{\circ}$. The inorganic LaCl$_{3}$(Ce) scintillator with a 2-cm thick $^{6}$Li polyethylene shield is also placed at 135$^{\circ}$ upstream.}
    \label{fig:Experimental setup}
\end{figure*}

The hybrid setup for ($n$,$\gamma$) cross section measurements, first used in the 2022 campaign, is made of an array of nine sTED units, two conventional large volume C$_{6}$D$_{6}$, and one LaCl$_{3}$(Ce) detectors at different angles and distances as shown in Fig.~\ref{fig:Experimental setup}. The segmented total-energy detector sTED~\cite{Alcayne:2023,Alcayne:2024} was primarily designed for coping with the extreme count rates characteristic of the high neutron flux at n\_TOF EAR2. The active volume of each detection unit, aluminum housed, is approximately one-ninth of a liter of C$_{6}$D$_{6}$ liquid scintillator~\cite{Mastinu:2013}. In this way, a compact block of 3$\times$3 detector modules reaches the overall efficiency of a large-volume C$_{6}$D$_{6}$ detector (see Fig.3 in Ref.~\cite{Alcayne:2023}). Compared to them, sTED modules also incorporate Photo-Multiplier Tubes (PMTs) Hamamatsu-R11265U, which are designed to withstand high counting rates and provide a fast time response. The use of this PMT helps to mitigate the impact of the $\gamma$-flash. The latter is a prompt and intense flux of $\gamma$-rays and relativistic particles produced by the arrival of the proton bunch at the spallation target~\cite{Mengoni:2019}. This intense flux of particles can easily saturate the response of most radiation detectors in the vicinity of the beam-line (see Sec.~\ref{Sec:Stilbene}). Finally, this type of PMT also helps to reduce dead-time and pile-up effects (see Ref.~\cite{Balibrea:2024} for details). 

Once the limiting effect of the high count rate was improved thanks to the detector segmentation, the next step was to enhance the detection sensitivity by optimizing the $s/b$ ratio as introduced in Sec.~\ref{Sec:Stat}. After a systematic exploration of the background conditions in the neighbourhood region of the sample, it became clear that the maximum sensitivity is achieved for as close as possible and equidistant positions of the individual sTED modules with respect to the sample. Thus, instead of using sTED as a compact-block array as proposed in Ref.~\cite{Alcayne:2024}, the individual small-volume detectors were separated and set-up in a cylindrical geometry around the capture sample~\cite{BalibreaEPJWoC:2023,Lerendegui:2023,Domingo-Pardo:2023} as shown in both panels of Fig.~\ref{fig:Experimental setup}. In this configuration detectors avoid any direct interaction with the neutron beam which is important for suppressing undesired neutron interactions with the active volume. The sample under study is placed at the geometrical center of the setup, which coincides with the beam axis, and the distance of detectors from this axis is 4.5 cm, as short as possible to accomodate 9 sTED units.

The inorganic LaCl$_3$(Ce) detector with a size of 50$\times$50$\times$25 mm$^{3}$ was placed at an angle of 135$^{\circ}$ with respect to the beam direction and a distance of 8.5 cm from the sample position. To mitigate neutron sensitivity effects in the LaCl$_{3}$(Ce) detector a 2 cm thick $^{6}$Li-enriched high density polyethylene block, with a size of 5$\times$5 cm$^2$, was placed in the front of the face of the crystal. The two $1L$ volume C$_{6}$D$_{6}$ detectors were also placed at an angle of 135$^{\circ}$ and a distance of 17~cm from the sample position. The larger distances for all these detectors were needed to compensate their large intrinsic and geometric efficiency in such a way that total count rates per detector were still tolerable and comparable to those of the individual sTED modules.
These large-volume detectors enable one to control different important experimental aspects, such as angular distribution effects or details of the capture cascade. In turn, the proper characterization of such effects are vitally important for an accurate assessment of several yield correction factors related to the measuring technique~\cite{Balibrea:2024}.

All detectors were calibrated using standard $\gamma$-ray calibration sources, including $^{137}$Cs, $^{207}$Bi, $^{60}$Co, and Am$/$Be, covering $\gamma$ rays with energy up to $\sim$4.4 MeV. A common threshold of 200 keV was applied in the data analysis for all detectors. The conversion between ToF and neutron energy was calculated using the relativistic formula~\cite{SCHILLEBEECKX:2012} with a fixed distance of 19.8 m.

In order to illustrate the sensitivity performance of the sTED compared to the large-volume detectors, a short test using $^{93}$Nb($n,\gamma$) was performed before the  $^{94}$Nb($n,\gamma$) cross section measurement campaign~\cite{BalibreaCDS:2020} using the experimental setup shown in Fig~\ref{fig:Experimental setup}. A full paper with the analysis of both cross sections will be presented in a dedicated publication. 

Two different sample configurations were measured. The first one corresponded to $^{93}$Nb sample in the form of a disc with 2 mm thickness and 16.34 mm diameter mounted on a 70 $\mu$m thick Kapton backing. The second configuration, devoted to background estimation, consisted only of 70 $\mu$m thick Kapton backing. These configurations are labelled as {\it{test}} and {\it{background}}, respectively. The number of protons devoted to each configuration, that in turn are proportional to the neutron beam intensity~\cite{Mengoni:2019}, was $N^{\prime} = 1.82\cdot$10$^{16}$ and $N = 8.77\cdot$10$^{16}$, respectively. This measurement was used to evaluate the $s/b$ performance or $p$-value from Eq.~\ref{eq:S1} for each individual detection system.

\begin{figure*}
    \centering
    \begin{tabular}{c c}
    \includegraphics[width=0.5\columnwidth]{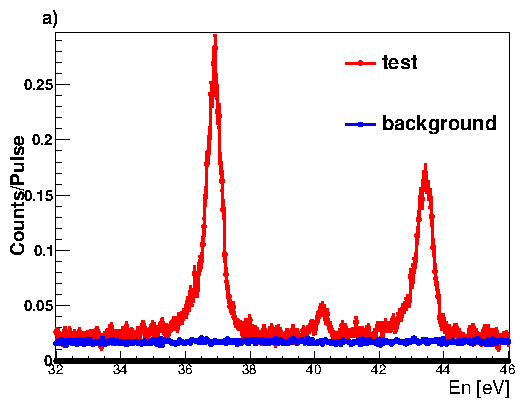} &
    \includegraphics[width=0.5\columnwidth]{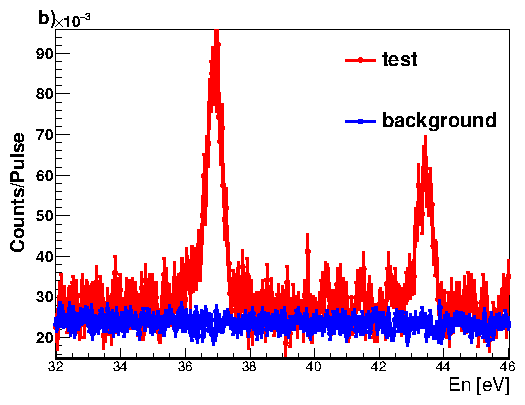} \\
    \includegraphics[width=0.5\columnwidth]{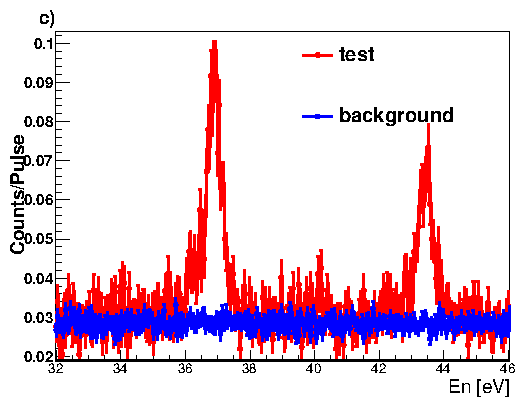} &
    \includegraphics[width=0.5\columnwidth]{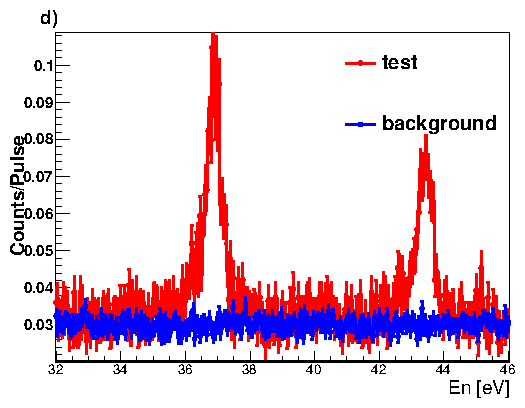}
    \end{tabular}
    \caption{Neutron energy spectra registered by all detection systems during the experiment in the neutron energy range corresponding with the two first neutron resonances of $^{93}$Nb. Panels a), b), c) and d) corresponds to sTED, LaCl$_{3}$(Ce) and both C$_{6}$D$_{6}$ $\gamma$-ray detection systems, respectively.}
    \label{fig:Example_resonances}
\end{figure*}

The measured count rate spectra normalized to the nominal proton-beam intensity, as a function of neutron energy, are shown in Fig.~\ref{fig:Example_resonances} for the sTED array (a), the LaCl$_{3}$(Ce) detector (b), and the two large-volume C$_{6}$D$_{6}$ detectors in panel (c) and (d). The {\it{test}} and {\it{background}} {\it{configurations}} are displayed with red and blue curves, respectively. The neutron-energy range shown in Fig.~\ref{fig:Example_resonances} covers the first two neutron resonances in $^{93}$Nb at 37 and 43 eV, which are well known from previous works~\cite{Drindak:2006,Wang:2011,Endo:2022}. 
The (beam-off) background related to the environmental activity is negligible in all cases and will not be discussed further.

The results obtained for the {\it{background configuration}} in all the detection devices (see Fig.~\ref{fig:Example_resonances}) indicate that the background level is similar in the neighborhood of the sample position, regardless of the exact detector position or distance, at least within the region of about 20~cm around the beam axis. On the other hand, as expected, the {\it{test configuration}} shows a much larger sensitivity to the neutron resonances for the sTED modules than for other detection systems. Indeed, for the allocated $N^{\prime}$ in the {\it test configuration}, the two prominent resonances in $^{93}$Nb+n are clearly visible in all detection systems. However, only the sTED array (panel a) allows one to visually identify the small resonance at 40 eV due to a Ta-contaminant in the sample.

\begin{figure}
    \centering
    \includegraphics[width=0.5\columnwidth]{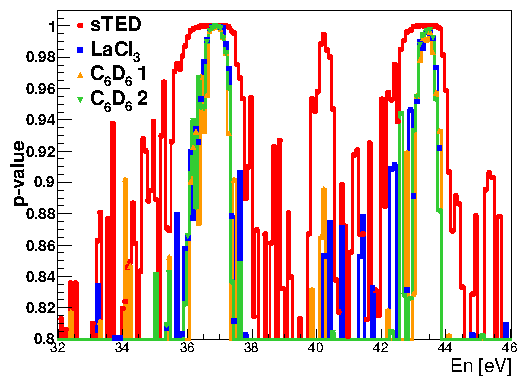}
    \caption{The significance of a signal observation, expressed as $p$-value, as a function of the neutron energy for the different experimental devices.}
    \label{fig:Significances}
\end{figure}

The larger sensitivity of the sTED detectors is to be ascribed to the higher $s/b$ ratio, which in turn arises from the inverse-squared sample-to-detector distance dependence of the signal intensity and the approximately constant background level in the sample surroundings. This fact reflect that it is much better a geometric optimization than the use of large active volumes. The relative $s/b$ ratio with respect to sTED in the neutron energy range from 32~eV to 46~eV is approximately 0.24 for the C$_{6}$D$_{6}$ detectors and 0.35 for the LaCl$_{3}$(Ce) detector. This magnitude was calculated using the neutron resonance integrals and background shown in Fig.~\ref{fig:Example_resonances}. The statistical significance of the results for the different detection systems was calculated using the $p$-value. The latter indicates the probability of rejecting the hypothesis of background-only, as given by Eq.~\ref{eq:p-value} and Eq.~\ref{eq:S1}. In other words, a $p$-value close to unity indicates a strong confidence that the measured data corresponds to a resonance, and not to a casual background fluctuation. The calculation was performed as a function of the neutron energy using individual neutron energy bin. The results are shown in Fig.~\ref{fig:Significances} (note the cut-off value at 0.8). Using a $p$-value threshold of 0.95, both the large-volume C$_{6}$D$_{6}$ and LaCl$_{3}$(Ce) detectors detect a signal different from the background around the resonance energies: 36.5-37.5 eV and 43.0-44.0 eV. However, the $p$-values above-threshold for sTED appear in a significantly wider energy range around the resonances, 35.5-38 and 42.5-44.5 eV, as well as around the resonance at 40 eV (a Ta contaminant), which is not detected with the other systems.

In short, the results depicted in Fig.~\ref{fig:Example_resonances} and Fig.~\ref{fig:Significances} indicate that the measuring time needed to achieve a statistical significance comparable to the sTED array with both C$_{6}$D$_{6}$ and LaCl$_{3}$(Ce) would be about 3 times larger. Thus, the sTED array in this geometric configuration can measure ($n$,$\gamma$) cross-sections approximately three times smaller than those measurable by other detection systems, enabling the study of more challenging cross sections at CERN n\_TOF EAR2, such as $s$-process bottlenecks, branching points, or samples with a reduced number of atoms~\cite{Domingo-Pardo:2023}. Apart from detection sensitivity or signal-to-background ratio, another aspect that becomes of importance for a capture experiment is the capture-cascade detection efficiency. This can be best studied from a Monte Carlo perspective, as discussed in the following section.

\section{Monte Carlo simulation of the hybrid ($n$,$\gamma$) setup at n\_TOF EAR2}\label{Sec:MC}

\begin{figure*}
    \centering
        \includegraphics[width=1.0\columnwidth]{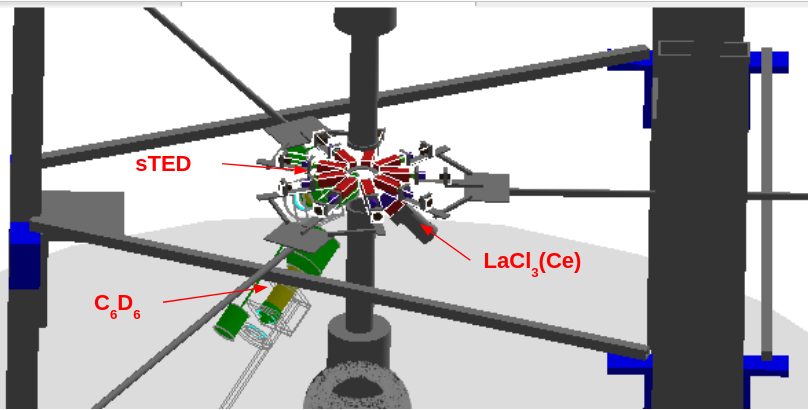}
    \caption{Detailed geometry of the experimental setup introduced in Sec.~\ref{Sec:PoC} implemented in \textsc{Geant4} simulations. The positions and distances of the $\gamma$-ray detection device models correspond to the experiment.}
    \label{fig:MC_setup}
\end{figure*}

A Monte Carlo simulation of the hybrid experimental setup was created using the \textsc{Geant4} toolkit (version 4.10.7)~\cite{Allison:2016} with the aim of comparing the different detection systems in terms of efficiency. The implemented geometry is illustrated in Fig.~\ref{fig:MC_setup}, which can be compared directly with the pictures of the actual experimental setup in both panels of Fig.~\ref{fig:Experimental setup}. The MC simulation includes all detection devices along with their housing materials, quartz optical windows and PMTs at their corresponding experimental positions. Additionally, aluminum and carbon fiber detector holding structures, neutron beam pipes, and the three pillars anchoring the experimental setup present in EAR2 were incorporated into the geometry to make it as close as possible to the reality. For the MC simulations, a modular physics approach was employed to automatically handle different categories of physics, such as electromagnetic (EM), hadronic, and decay processes. In the current code, the modular physics setup includes the standard electromagnetic package option 4, radioactive decay, the Particle High Precision model, and neutron thermal scattering. This MC application, therefore, enables accurate simulations of both $\gamma$-ray and neutron interactions across a broad range of energies.

\begin{figure}
    \centering
    \includegraphics[width=0.5\columnwidth]{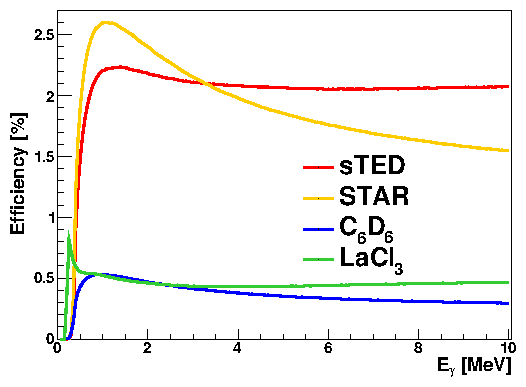}
    \caption{Calculated MC detection efficiency as a function of incident $\gamma$-ray energy for sTED (red), STAR (Orange), individual C$_{6}$D$_{6}$ (blue) and LaCl$_{3}$(Ce) detectors (green).}
    \label{fig:GammaEfficiency_Stilbene}
\end{figure}

The $\gamma$-ray detection efficiency curve for each detector was determined by simulating mono-energetic $\gamma$-rays isotropically emitted from the sample position assuming a point like source. A common threshold of 200~keV in deposited energy was applied to all detectors similarly to experimental values (see Sec.~\ref{Sec:PoC}). The efficiency as a function of the $\gamma$-ray energy for the sTED array, the individual large-volume C$_{6}$D$_{6}$ detectors, and the LaCl$_{3}$(Ce) detector are displayed in Fig.~\ref{fig:GammaEfficiency_Stilbene}. 

Because of the different sample-detector distances at which each detection system can operate in the high-flux conditions of EAR2 (see Sec.~\ref{Sec:PoC}), there are notable differences in absolute detection efficiency. The $\gamma$-ray efficiency of the sTED array is a factor between 4 and 7 higher than that of a C$_{6}$D$_{6}$ detector. A similar ratio is found with respect to the LaCl$_{3}$(Ce) detector. This fact, together with the experimental results discussed in Sec.~\ref{Sec:PoC}, reflects the twofold advantage of using small-detection volumes in high neutron flux environments. On the one hand, detector-to-sample distances can be significantly shortened (Fig.~\ref{fig:Experimental setup}) thus enhancing $s/b$ ratio (Fig.~\ref{fig:Significances}), as well as overall detection efficiency (Fig.~\ref{fig:GammaEfficiency_Stilbene}). On the other hand, a further degree of flexibility in detector geometry is achieved (see Fig.~\ref{fig:Experimental setup}), which allows for the additional steps in terms of sensitivity optimization with commonly used scintillation materials. It is noteworthy that further refinements and optimizations may be also attained by means of new scintillation materials, as it discussed in the following section.

\section{Future prospects based on solid deuterated-stilbene scintillators}\label{Sec:Stilbene}

The possibility to utilize solid plastic scintillators for measuring radiative neutron-capture cross sections via the ToF technique has been a topic of discussion for many years. However, the presence of hydrogen in significant amounts in commercially available scintillation materials, like BC-537, has hindered this possibility owing to the high neutron sensitivity and related backgrounds. In the last decade, developments of new solution-growth methodologies for high-quality conventional trans-stilbene organic crystals allowed to prepare high purity deuterated stilbene crystals (stilbene-d12) \cite{Becchetti:2017,Carman:2018}. Their chemical composition (C$_{10}$D$_{12}$), high density (1.24~g/cm$^{3}$), large $n$/$\gamma$ pulse-shape discrimination capabilities, large light yield output, and smaller or negligible chemical risks, at variance with purified deuterated benzenes, makes this scintillation material a very promising replacement of liquid C$_{6}$D$_{6}$ organic scintillators, which have been extensively used for neutron-capture ToF experiments over the last decades.

There are other potential benefits of using solid scintillators. Specifically, active detection volumes in solid (and non hygroscopic) form allow one to use a thinner encapsulation, avoid expansion volumes required with liquid-based detectors, and permits also to eliminate the boro-silicate quartz window, commonly used for optical coupling between the PMT and the liquid scintillator. Finally, a solid scintillator can be more easily coupled to a SiPM photosensor, thereby reducing further the amount of materials in the detection device~\cite{Balibrea:2021}. All these aspects should further contribute to a reduction of the intrinsic neutron sensitivity of the detection device, thus enabling improved accuracy for the measurement in neutron energy regions with a dominant scattering cross section~\cite{Kappeler:2011,Plag:2003}. 

Empowered by this motivation, we present here a conceptual design study for a future Stilbene-d12 deTector ARray, referred to as STAR, primarily aimed at ($n,\gamma$) measurements of very small and/or radioactive samples at CERN n\_TOF EAR2. First experimental results with one STAR-module prototype will then presented and discussed in the following section.

\begin{figure}
    \centering
\includegraphics[width=0.5\columnwidth]{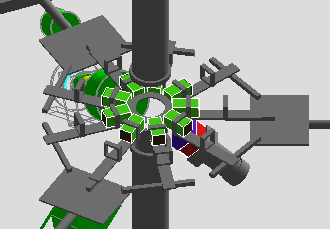}
    \caption{Graphical representation of the STAR experimental setup as implemented in \textsc{Geant4}. sTED detectors in Fig.~\ref{fig:MC_setup} have been replaced by STAR detectors, while the rest of elements, including the structural materials are placed in the same position as for the setup introduced in Sec.~\ref{Sec:PoC}.}
    \label{fig:MC_setup_Stilbene}
\end{figure}

For the design and optimization of STAR we conducted a MC study using a modified \textsc{Geant4} implementation of the hybrid setup introduced in Sec.~\ref{Sec:MC}. The modification consisted of replacing the individual sTED detectors with nine stilbene-d12 detectors. Each individual stilbene crystal has a size of 25$\times$25$\times$50 mm$^{3}$, resulting in a total active volume of 281.25 cm$^{3}$. The distance of the face of the individual crystals from the center of the setup was kept the same as for sTED. In this way, one avoids any possible direct interaction with the neutron beam and their performance can be directly compared to other detectors in the hybrid setup. Considering STAR dimensions, this represents a reduction factor of 3.56 in volume compared to sTED while preserving a comparable $\gamma$-ray efficiency. Further, in the simulations a 1.5 mm thick SiPM was included instead of a PMT used for sTED, matching the small 25$\times$25 mm$^{2}$ detector face. The active volume of the individual modules was housed using 1 mm Teflon wrap combined with 0.5~mm aluminum layer. The geometry implementation in the \textsc{Geant4} simulation is displayed in Fig.~\ref{fig:MC_setup_Stilbene}. 

The $\gamma$-ray detection efficiency for STAR was obtained in a similar fashion as described in Sec.~\ref{Sec:MC} and thus, a direct comparison with the rest of the detection systems is straight forward and shown in Fig.~\ref{fig:GammaEfficiency_Stilbene}. Compared to the large-volume C$_{6}$D$_{6}$ and LaCl$_{3}$(Ce) detectors the efficiency of STAR is larger by a factor $\approx$ 4-6$\times$, depending on $\gamma$-ray energy. Detection efficiencies of STAR and sTED are thus similar in magnitude but display some difference in the $E_{\gamma}$ dependence. Although the efficiency of STAR is lower than for sTED for a significant range of $\gamma$-ray energy dependence, the total efficiency for actual cascades could be significantly higher, especially for heavy nuclei, where the $\gamma$-ray de-excitation is dominantly made via several transitions rather by a direct transition to the ground state.

\begin{figure}
    \centering
    \includegraphics[width=0.5\columnwidth]{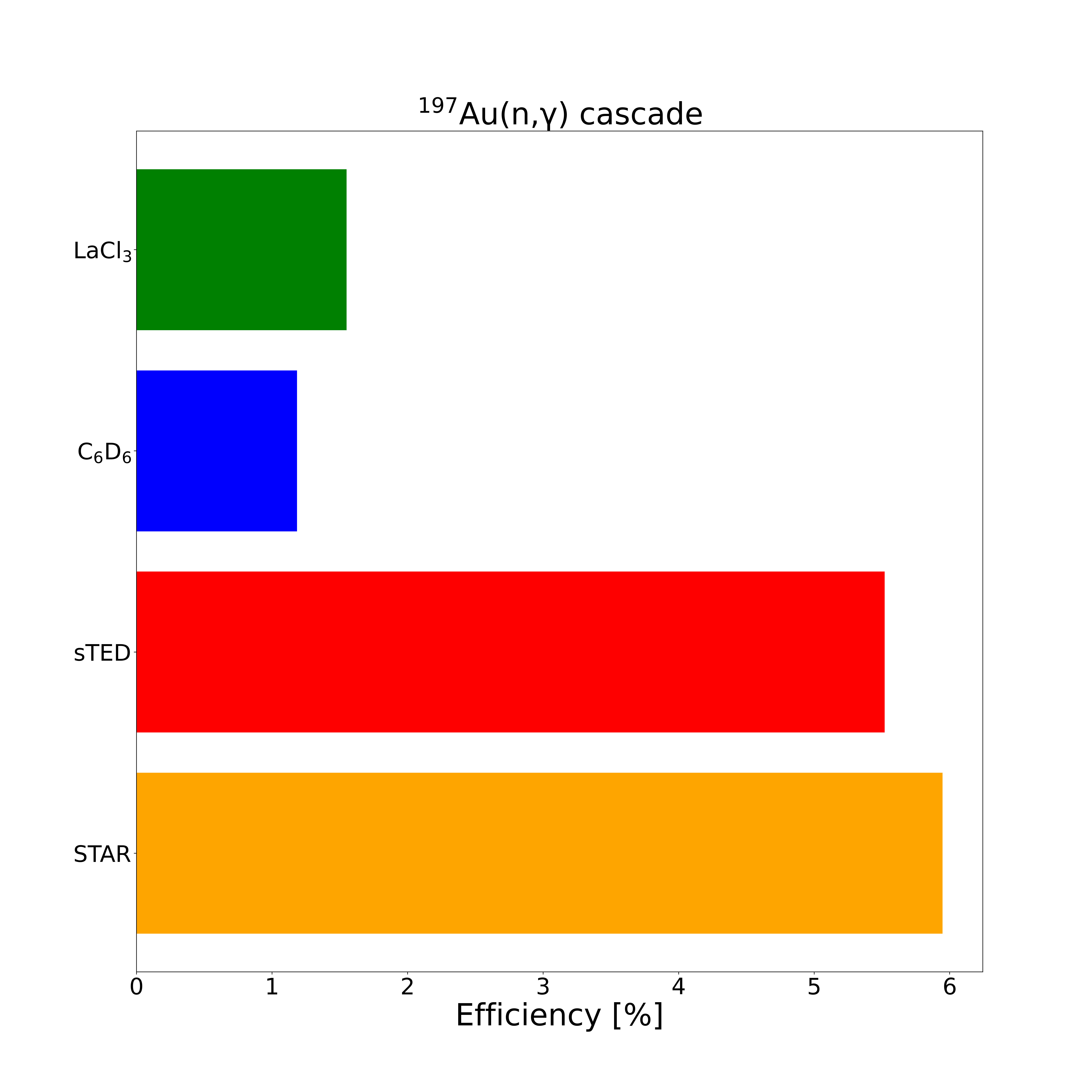}
    \caption{Calculated MC detection efficiency for $^{197}$Au($n$,$\gamma$) decay for sTED (red), STAR (Orange), individual large-volume C$_{6}$D$_{6}$ (blue) and LaCl$_{3}$(Ce) detectors (green).}
    \label{fig:AuEfficiency}
\end{figure}

In order to illustrate this feature, a simulation of the response of individual detector systems was undertaken for s-wave resonances formed in $^{197}$Au($n$,$\gamma$) reaction. This reaction is commonly used for yield normalization purposes in neutron-capture ToF experiments~\cite{SCHILLEBEECKX:2012,MacKlin:1969,Massimi:2011}. The $\gamma$-ray cascades for $^{197}$Au($n$,$\gamma$) reaction were simulated with \textit{DICEBOX}~\cite{Beckvar:1998} code. These cascades give a very good reproduction of spectra from C$_{6}$D$_{6}$ detectors measured at n\_TOF~\cite{Balibrea:2024} as well as from DANCE detector~\cite{HEIL:2001,REIFARTH:2004}. The capture efficiency obtained from this simulation for each detection system is displayed in Fig.~\ref{fig:AuEfficiency}. In summary, the capture-cascade detection efficiency for $^{197}$Au($n$,$\gamma$) is expected to be about 10\% larger for STAR compared to sTED because of the $\gamma$-ray cascades pattern. Similar variations, in one direction or another, are expected depending on the exact $\gamma$-ray energies emitted during the decay of individual isotope.

\section{First experimental results with a stilbene-d12 prototype}\label{sec:ExpStilbene}

A first STAR module prototype was assembled, consisting of a small 25$\times$25$\times$13~mm$^{3}$ stilbene-d12 crystal produced at Lawrence Livermore National Laboratory~\cite{Becchetti:2017,Carman:2018}. The crystal was wrapped with 0.5~mm Teflon and 0.5~mm adhesive black tape. The module was optically coupled to a Hamamatsu-R11265U PMT and tested at the Instituto de Física Corpuscular (IFIC) and later also at CERN n\_TOF EAR2~\cite{BalibreaCDS:2023}. The goal of the present prototype and measurements was twofold. On the one hand, these measurements were intended to study and characterize the neutron-gamma Pulse Shape Discrimination (PSD) capabilities of the stilbene-d12 crystal coupled to the aforementioned PMT. On the other hand, the measurements enabled one to assess the performance of this small prototype in a representative ToF experiment, thereby comparing the stilbene-d12 response with respect to large-volume C$_{6}$D$_{6}$ detectors. 
 
\subsection{Machine-learning aided neutron-gamma pulse-shape discrimination with stilbene-d12}

The PSD capability of stilbene-d12 was investigated at IFIC using a Lecroy Oscilloscope (64MXi-A), equipped with a bandwidth of 600~MHz and a maximum sampling rate of 10 Gs/s. Two radioactive sources of $^{22}$Na and $^{252}$Cf were utilized for this aim. The first source was used for energy calibration of the detector and to cross-check the $\gamma$-ray pulse shape waveform. The second source, which emits both neutrons and $\gamma$-rays, was used to perform pulse shape discrimination for the detector. 

The raw detector pulses were digitized at 5 Gs/s and recorded for offline analysis. Before the analysis all registered waveforms were first aligned in time using a constant-fraction digital algorithm. The signal baseline level, calculated as the average of 100 samples before signal, was also subtracted and all the acquired waveforms. All the treated digitized waveforms were normalized in amplitude. 

Two different PSD methodologies were investigated, namely, the charge comparison method~\cite{SENOVILLE:2020} and the unsupervised Machine Learning $k$-means++~\cite{Lloyd:1957,MacQueen:1967,Bishop:2006} algorithm implemented in the \textsc{scikit-learn} library~\cite{scikit-learn}. For the charge comparison method, a PSD parameter, $Q$, is defined as the ratio between the fast ($I_{fast}$) and the total ($I_{total}$) waveform integral ($Q=I_{fast}/I_{total}$)~\cite{SENOVILLE:2020}. An algorithm for determination of integration time windows was implemented in order to obtain the best value for the Figure of Merit (\textit{FoM})~\cite{SENOVILLE:2020} defined as

\begin{equation}
FoM=\frac{D_{n}-D_{\gamma}}{W_{n}+W_{\gamma}}
\end{equation}
where $D_{n}$ and $D_{\gamma}$ are the mean positions of the neutron and $\gamma$-ray $Q$ parameter, respectively, and $W_{n}$ and $W_{\gamma}$ are the corresponding \textsc{fwhm}. It was found that the best $FoM$ was reached for a fast- and total-time integration windows of 20~ns and 100~ns, respectively. Using a minimum deposited energy of 0.2 MeVee one obtains a $FoM$ of 1.17. This result is comparable with the results reported previously~\cite{Becchetti:2017,Carman:2018}. Therefore, one can conclude that the use of the fast PMT is not affecting noticeably the PSD capability when compared with previous works.

\begin{figure}[!htbp]
    \centering
    \includegraphics[width=0.5\columnwidth]{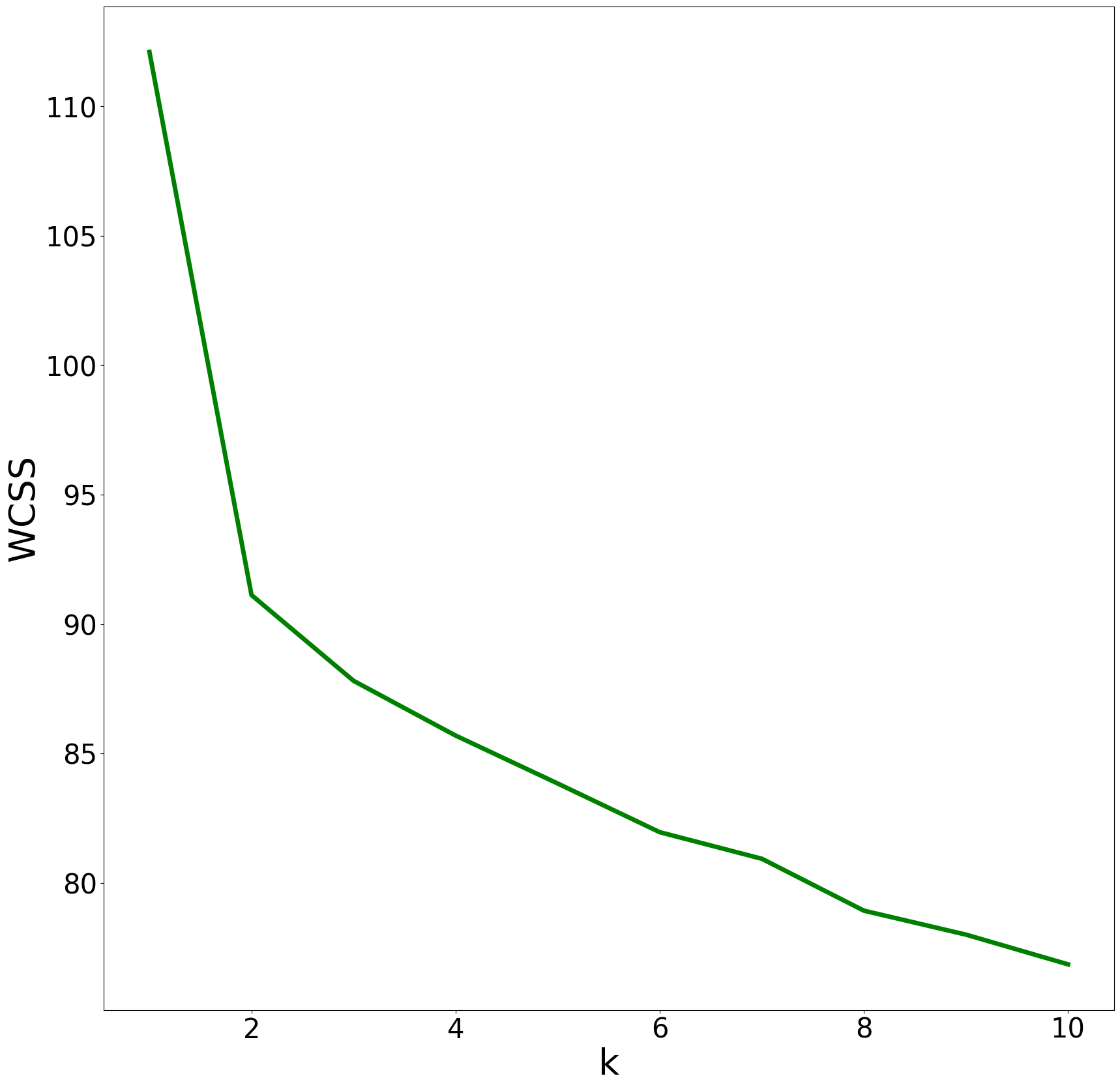}
    \caption{WCSS values as a function of number of clusters, $k$. The change in slope determines the optimum number of clusters for the $k$-mean++ algorithm.}
    \label{fig:Elbow_methodology}
\end{figure}

For the unsupervised Machine Learning $k$-means++ algorithm, the set of waveforms $(\vec{x_{1}},...,\vec{x_{n}})$, was fed into the algorithm aiming to partition the $^{252}$Cf dataset into $k$ clusters. Each waveform is a $d$-dimensional vector and therefore, the clustering algorithm will work in a $d$-dimensional space, corresponding to the length of the digitized waveforms. The algorithm minimizes the within-cluster sum of squares (WCSS) defined as
\begin{equation}
    \mathrm{argmin}_{S} \sum^{k}_{i=1}\sum_{\vec{x}\in S_{i}}{\left|\left|\vec{x}-\vec{\mu_{i}}\right|\right|^{2}}
\end{equation}
where $\vec{\mu_{i}}$ is the $d$-dimensional mean of points within the cluster $S_{i}$. The number of clusters, $k$, is the only parameter to be optimized during the process and it is usually determined with the "elbow" technique~\cite{Thorndike:1953}. It is a heuristic method consisting of plotting WCSS as a function of $k$ as shown in Fig.~\ref{fig:Elbow_methodology} and identifying the "elbow" point where WCSS changes the slope abruptly.
In this particular case, approximately $10^{5}$ waveforms were fed to the algorithm. Each of them had a fixed length of 350 samples covering the range of 70 ns due to memory constrains of the computer used for this work. Fig.~\ref{fig:Elbow_methodology} clearly indicates that the curve slope changes drastically at $k$=2. In fact, 2 clusters correspond to a physical solution of the problem since there are only two types of particles interacting with the active volume of the detector: $\gamma$-rays and neutrons emitted from the $^{252}$Cf calibration source. 

\begin{figure}
\centering
\includegraphics[width=0.5\columnwidth]{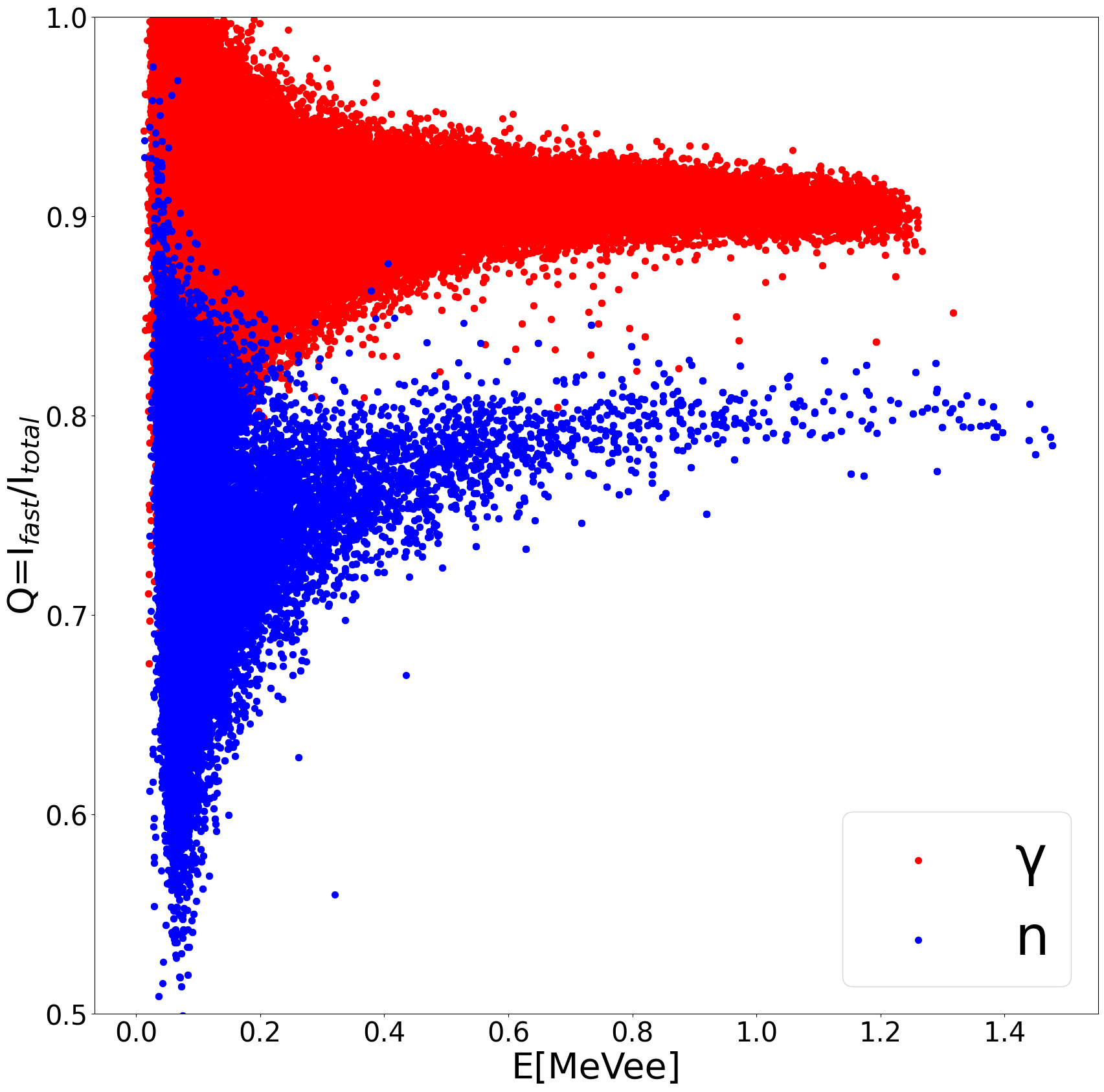}
\caption{PSD parameter $Q$ derived from charge comparison method as a function of the the deposited electron equivalent energy for the small stilbene-d12 module for pulses from the $^{252}$Cf. Blue and red dot points indicate neutron and $\gamma$ signals as derived from $k$-means++ algorithm, respectively.}
\label{fig:PSD_ML}
\end{figure}

In order to compare with the traditional charge comparison PSD $Q$ parameter, the corresponding $k$-means++ derived label ($\gamma$ or neutron) is plotted together with the $Q$ parameter as a function of the deposited electron equivalent energy in Fig~\ref{fig:PSD_ML}. The labels obtained from the $k$-means++ are presented as red and blue markers, corresponding to $\gamma$- and neutron-tagged events, respectively. Identification of $\gamma$-rays and neutrons was based on signals from $^{22}$Na.  

The $k$-means++ results are in good agreement with the charge comparison method. The distribution of points from $k$-means++ looks very reasonable and indicates a possibility to extend particle identification further down in electron equivalent deposited energy. This possibility will be investigated in future works. Such development might provide better discrimination of neutron from $\gamma$-ray particles.

\begin{figure}
\centering
\includegraphics[width=0.5\columnwidth]{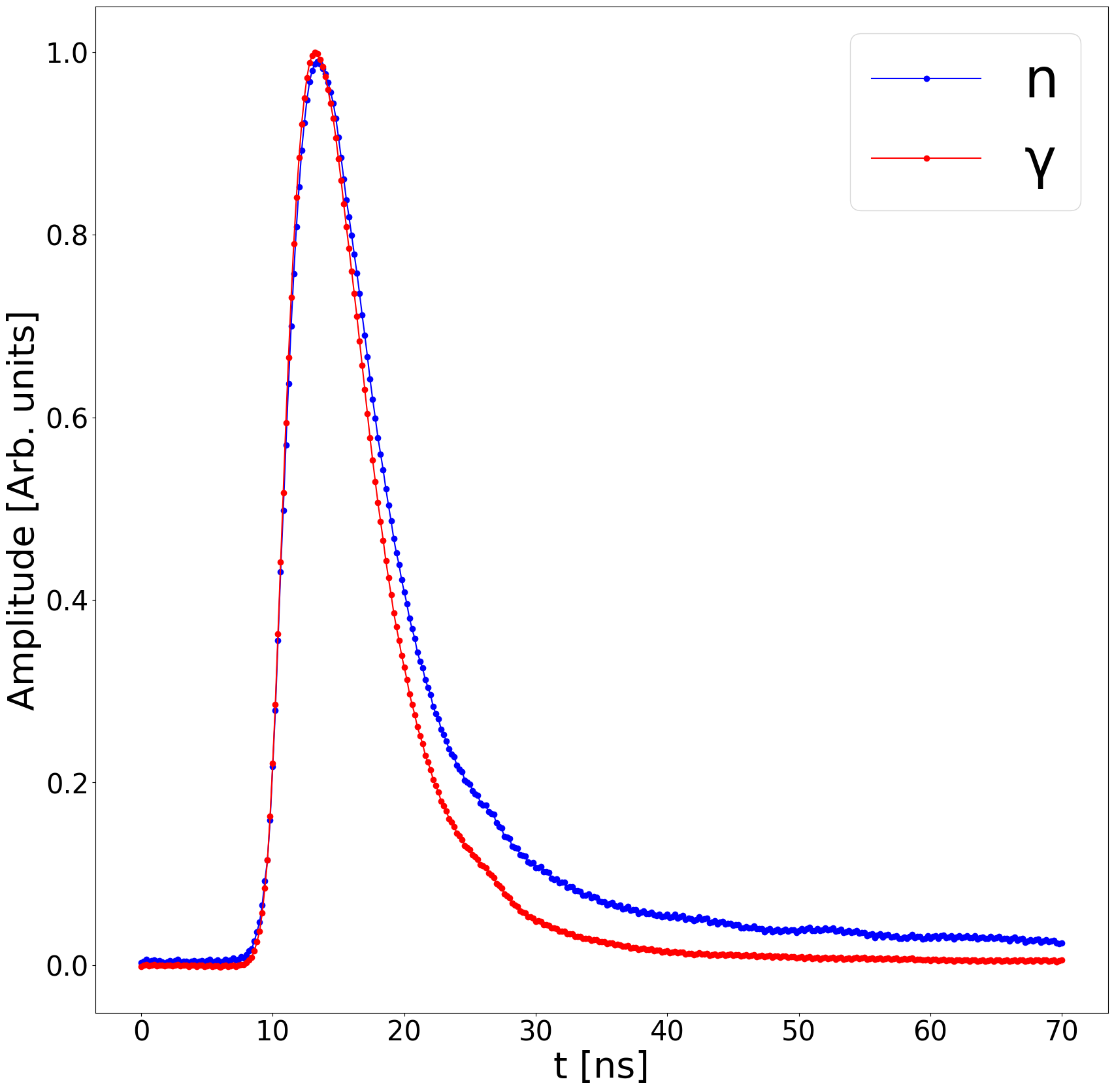}
\caption{Average neutron and $\gamma$-ray events waveforms calculated based on the identification obtained from the $k$-means++ approach.}
\label{fig:Pulse_shape_ML}
\end{figure}

For completeness, Fig.~\ref{fig:Pulse_shape_ML} shows the average pulse shape experimentally obtained by means of the $k$-means++ approach. The waveforms, with an average pulse time width of only 18~ns and 22~ns \textsc{fwhm} for $\gamma$ and neutron signals are well suited for high count rate environments, such as the one at n\_TOF EAR2~\cite{Balibrea:2024}.

\subsection{Characterization measurements at CERN n\_ToF EAR2}\label{sec:Exp_stilbene}

\begin{figure}
\centering
\includegraphics[width=0.5\columnwidth]{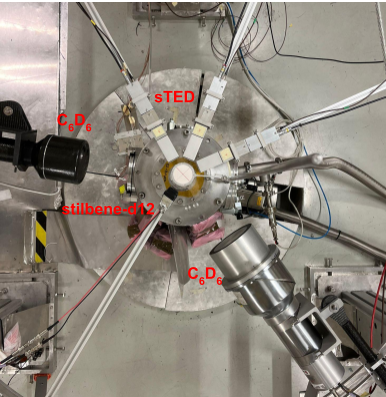}
\caption{Top view of the experimental setup used during the test performed at n\_TOF EAR2. Two large-volume C$_{6}$D$_{6}$ detectors, three sTED modules and the stilbene-d12 detector were placed at 90$^{\circ}$ and different distances (see text) to the center of the neutron beam line.}
\label{Fig:ExpSetupTest}
\end{figure}

The small STAR module prototype was then tested in a dedicated experimental campaign at n\_TOF EAR2 together with several other detectors~\cite{BalibreaCDS:2023}. An upstream picture of the experimental setup is shown in Fig~\ref{Fig:ExpSetupTest}. Two large-volume C$_{6}$D$_{6}$ detectors, and three sTED individual modules were placed together with the stilbene-d12 prototype, all of them at 90$^{\circ}$ with respect to the neutron beam. The distance from each detector to the sample (neutron-beam axis) was adjusted in order to achieve similar count rates in all detectors. All detectors were fully functional even under the largest count-rate conditions reached and the corresponding data could be reliably analyzed and compared. In particular, the large-volume C$_{6}$D$_{6}$ were placed at 17~cm from the sample, whereas the sTED modules were located at 5~cm distance, a slightly different distance from the nominal one described in Sec.~\ref{Sec:PoC}. The stilbene-d12 prototype was placed at the same distance from the beam axis as individual sTED modules. 

\begin{figure}
\centering
\includegraphics[width=0.5\columnwidth]{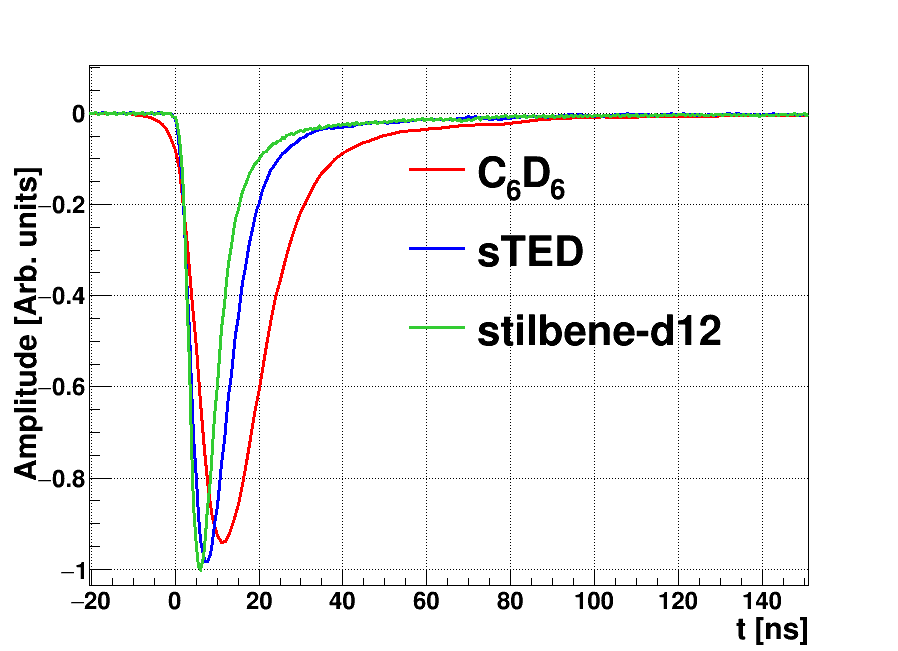}
\caption{Average pulse shape obtained for all the detectors present during the experimental campaign at n\_TOF EAR2 described in Sec.~\ref{sec:Exp_stilbene}.}
\label{Fig:PulseShape_nTOF}
\end{figure}

The time-response function of the stilbene-d12 prototype was compared with the other detectors. The main interest of this measurement was to investigate the time response of the STAR module in realistic conditions. Fig.~\ref{Fig:PulseShape_nTOF} shows the average pulse shape waveform of all different detectors obtained with an $^{88}$Y calibration source placed at the center of the experimental setup. 
The risetime, calculated as the time difference between 10\% and 90\% of the signal maximum is of 3~ns, 4~ns and 8~ns for the stilbene-d12, the sTED and C$_{6}$D$_{6}$ detectors, respectively. The pulse time-width (\textsc{fwhm}) for large-volume C$_{6}$D$_{6}$, sTED and stilbene-d12 are 40~ns, 23~ns and 18~ns, respectively. To some extent, the better timing characteristics of the sTED and stilbene-d12 detectors are to be ascribed to their smaller volume and consequently more efficient light collection~\cite{Balibrea:2024}. Therefore, slightly worse timing characteristics may be found for the future (larger size) stilbene-d12 crystals of 25$\times$25$\times$50~mm$^3$ planned to be used in STAR (see Sec.\ref{Sec:Stilbene}). However, the excellent results found here lend confidence that the deuterated stilbene will provide also sufficiently good time-response for neutron-capture experiments in the high-flux conditions of EAR2. 

\begin{figure}
\centering
\includegraphics[width=0.5\columnwidth]{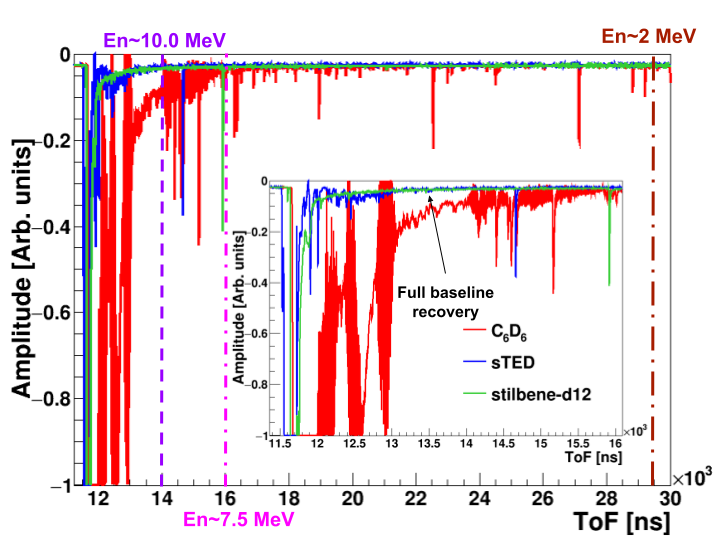}
\caption{Response of the different detectors to the prompt $\gamma$-flash with an empty sample. The inset shows a zoom in the high-neutron energy region.}
\label{Fig:Gflash_nTOF}
\end{figure}

Another important aspect for any neutron-capture detection setup at EAR2 is the recovery time needed after the prompt $\gamma$-flash. Fig.~\ref{Fig:Gflash_nTOF} illustrates the recovery from the $\gamma$-flash at EAR2 for C$_{6}$D$_{6}$, sTED and stilbene-d12 detectors, covering a ToF region down to a few MeV neutron energy. The figure corresponds to a signal obtained with no sample used, which corresponds to the minimum (ideal) $\gamma$-flash response. The large-volume C$_{6}$D$_{6}$ detectors recover the baseline level at times corresponding to energies of a few MeV($\sim$7~MeV). On the other hand, the response of the sTED and stilbene-d12 detectors reaches the baseline already at about three times higher neutron energies. This is an important feature of small-volume detectors that allows to measure at higher neutron energies than with the large-volume C$_{6}$D$_{6}$ ones.

\begin{figure}
\centering
\includegraphics[width=0.5\columnwidth]{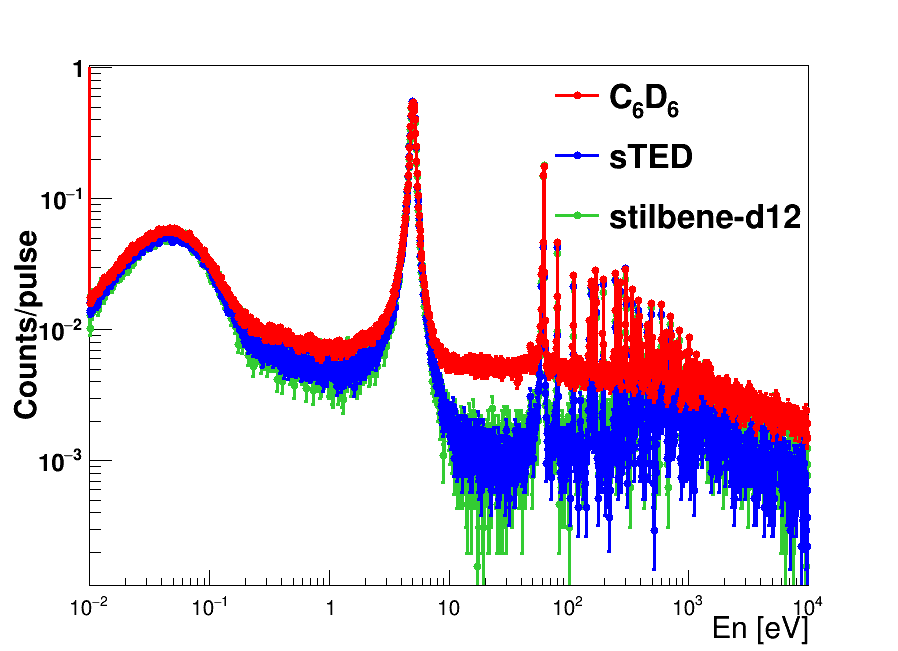}
\caption{Experimental yield of $^{197}$Au($n,\gamma$) measured at n\_TOF EAR2 with different detection systems normalized to the saturated resonance. Results are shown for a large-volume C$_{6}$D$_{6}$ detector, the sTED detectors present and the stilbene-d12 prototype described in Sec.~\ref{sec:Exp_stilbene}.}
\label{Fig:Au_yield_different_detectors}
\end{figure}

Finally, for a quick assessment of the signal-to-background ratio with the stilbene-d12 detector a short measurement with a gold sample was performed. Fig.~\ref{Fig:Au_yield_different_detectors} shows the measured count-rate spectra for all types of used detectors. For comparison purposes, the sTED and stilbene-d12 yields were normalized to C$_{6}$D$_{6}$ spectrum in the saturated region of the 4.9 eV gold resonance. The $s/b$ ratio for both sTED and stilbene-d12 is about a factor 9 better than for the large-volume C$_{6}$D$_{6}$ detector, when comparing the valley between 4.9 and 60 eV resonances. A detailed inspection of the spectrum measured with stilbene-d12 further revealed no contributions from chemical or isotopic contaminations arising from neutron-capture in the crystal itself. This result is very important and indicates the quality and purity of the stilbene-d12 crystal used in these tests.

In summary, the results presented above essentially validate the main features (timing performance, background, sensitivity) of the new stilbene-d12 material for neutron-capture time-of-flight experiments, and they pave the path for the future realization of the full STAR array. Further validation measurements are foreseen in the near future utilizing a STAR demonstrator based on stilbene-d12 crystals with a volume of 25$\times$25$\times$50~mm$^3$.

\section{Conclusions and outlook}\label{Sec:Conclusions}

Conventional C$_{6}$D$_{6}$ detectors regularly used for ($n$,$\gamma$) cross-section measurements over the past decades face significant limitations in performance when utilized in high-intensity Time-of-Flight measuring stations, such as EAR2 at the n\_TOF facility. These limitations arise as a consequence of their relatively large volume and readout photosensors. Experiments have shown that such effects can be mitigated by means of an array of small-volume detectors. An array of nine small C$_{6}$D$_{6}$ modules, known as sTED, represents nowadays the state of the art in this field. Initially designed as a compact block of 3$\times$3 sTED modules, in this work we have found that a cylindrical arrangement in a close geometry around the capture sample provides much better performance, particularly for optimizing the $s/b$ ratio, which is crucial for neutron-capture cross-section measurements involving small cross sections or samples with very few atoms.

One of the primary objectives of this paper is therefore to present the key features of the sTED setup in cylindrical configuration, derived from measurements using a hybrid configuration for ($n$,$\gamma$) experiments at CERN n\_TOF EAR2 facility in 2022. This setup includes nine sTED detector modules arranged around the capture sample, alongside with two large-volume conventional C$_{6}$D$_{6}$ detectors and a LaCl$_{3}$(Ce) detector positioned at 135$^{\circ}$ relative to the neutron beam. Using data from a short measurement of a $^{93}$Nb sample, we demonstrate a nearly tenfold improvement in the $s/b$ ratio with the sTED in cylindrical configuration compared to conventional C$_6$D$_6$ detectors. However, the contribution of regular C$_{6}$D$_{6}$ and a LaCl$_{3}$(Ce) detectors in the hybrid setup is still of relevance towards a better control of the systematic effects related to the angular distribution of $\gamma$-rays emitted in low-multiplicity capture cascades~\cite{Domingo:2006}, as well as other potential systematic effects~\cite{Abbondanno:2004}. The experimental results detailed in Sec.~\ref{Sec:PoC} demonstrate that the sTED array in this cylindrical configuration is sensitive to ($n$,$\gamma$) cross-sections approximately three times smaller than those detectable by other detection devices. Therefore, the use of this experimental setup will enable studies of more challenging cross-sections, such as $s$-process bottlenecks, branching points, or samples with a limited number of available atoms.

Apart from signal-to-background ratio, the efficiency for detecting capture events was explored by means of \textsc{Geant4} simulations. The results shown in Sec.\ref{Sec:MC} indicate that the efficiency of the sTED array for individual $\gamma$-rays is between 4 and 7 times higher than that of a C$_{6}$D$_{6}$ detector. In terms of capture-cascade detection efficiency, for the  $^{197}$Au($n$,$\gamma$) reaction (see Fig.\ref{fig:AuEfficiency}), the sTED efficiency was found to be 5 to 6 times higher than that of other detection systems.

The \textsc{Geant4} model was also utilized to explore alternative detection systems that could provide additional enhancements. Thus, in Sec.~\ref{Sec:Stilbene} a new detection system called STAR (Stilbene-d12 deTector ARray) was proposed and investigated both conceptually and experimentally. This next generation of neutron-capture detectors is expected to provide enhanced performance and improved chemical safety. The MC simulations indicate a comparable or slightly higher (10\%) efficiency compared to sTED for ($n$,$\gamma$) cross section measurements, as shown in Fig.\ref{fig:AuEfficiency}. However, one of the main advantages of STAR is expected to come from their smaller (solid) volume, and the larger flexibility in terms of detection-setup optimization and customization for each type of experiment.

To demonstrate the feasibility of the STAR detection setup, we conducted initial measurements using a small stilbene-d12 prototype of 25$\times$25$\times$13~mm$^{3}$ coupled to a fast PMT. The output signals have a width of 18 ns and 22 ns \textsc{fwhm} for $\gamma$-ray and neutron events, respectively. As shown in Sec.~\ref{sec:ExpStilbene}, the PSD value (1.17) is consistent with previous works. An unsupervised machine learning approach was developed to further explore the PSD at low deposited energies, yielding compatible results. In the future, this approach will be tested at n\_TOF to evaluate the possibility of automatic PSD as a means for further background rejection.

Some of the measurements for the small prototype were performed directly at EAR2, in a configuration that included additional detectors for comparison. As described in Sec~\ref{sec:Exp_stilbene}, the prototype has a pulse shape faster than the small-volume C$_6$D$_6$ detectors, fast recovery from the $\gamma$-flash in the MeV region and a signal-to-background ratio of a factor 9 larger compared with large-volume C$_{6}$D$_{6}$ detectors. In short, the key performance indicators of this small stilbene detector module yield great expectations for future experiments with STAR, involving high-intensity and high-quality neutron beams, both at CERN n\_TOF EAR2 and other high neutron flux facilities.

\section*{Acknowledgments}
The authors acknowledge support from the Spanish Ministerio de Ciencia e Innovaci\'on under grants PID2019-104714GB-C21, PID2022-138297NB-C21 and the financial support from MCIN, PCI2022-135037-2 funded by MCIN/AEI/10.13039/501100011033/ and the European Union NextGenerationEU and Generalitat Valenciana in the call PRTR PC I+D+i ASFAE/2022/027. This work was supported by European Union NextGeneration EU/PRTR project C17.I02.P02-SGI\_GICS Nuevas actuaciones en grandes infraestructuras de investigación europeas e internacionales, subproject C17.I02.P02.S01.S03 CSIC CERN. The corresponding author JB acknowledges support from grant ICJ220-045122-I funded by \\ MCIN/AEI/10.13039/501100011033. Author VB is a beneficiary of the Margarita Salas grant (MS21- 178) for the requalification of the Spanish university system from the Ministry of Universities of the Government of Spain, financed by the European Union, NextGenerationUE. Author JL acknowledges the support provided by postdoctoral grants FJC2020-044688-I by \\ MCIN/AEI/10.13039/501100011033 and CIAPOS/2022/020 funded by the Generalitat Valenciana and the European Social Fund and a PhD grant PRE2023 from CSIC. The work on stilbene-d12 development was performed under the auspices of the U.S. Department of Energy by Lawrence Livermore National Laboratory under Contract DE-AC52-07NA27344. Support from the funding agencies of all other participating institutes are also gratefully acknowledged.

\printcredits

\bibliographystyle{cas-model2-names}

\bibliography{Bibliography}



\end{document}